\documentclass[aps,prl,reprint,superscriptaddress,showpacs,nofootinbib]{revtex4-2}

\usepackage{hyperref}
\usepackage{amsmath}
\usepackage{amssymb}
\usepackage{graphicx}
\usepackage{color}
\usepackage{subfig}
\usepackage{bm}
\usepackage{gensymb}
\usepackage{bigints}
\usepackage[usenames,dvipsnames,svgnames,table]{xcolor}
\usepackage{pifont}
\usepackage{aas_macros}
\usepackage{orcidlink}
\usepackage{physics}

\usepackage[labelfont=bf]{caption}

\hypersetup{
    colorlinks=true,
    linkcolor=Blue,
    filecolor=Blue,
    urlcolor=MidnightBlue,
    citecolor=Blue,
   pdftitle={Cooling age}
}

\begin{document}

\title{\boldmath Cooling process of substellar objects in scalar-tensor gravity}

\author{Aleksander Kozak}
\affiliation{Institute of Theoretical physics, University of Wroclaw, pl. Maxa Borna 9, 50-206 Wroclaw, Poland,}
\email{aleksander.kozak@uwr.edu.pl}
\author{K\"{a}rt Soieva}
\affiliation{Laboratory of Theoretical Physics, Institute of Physics, University of Tartu,
W. Ostwaldi 1, 50411 Tartu, Estonia
}
\email{kart.soieva@ut.ee}
\author{Aneta Wojnar}
\affiliation{Departamento de F\'isica Te\'orica \& IPARCOS, Universidad Complutense de Madrid, E-28040, 
Madrid, Spain}
\email{awojnar@ucm.es}

\begin{abstract}
    Cooling processes of brown dwarf stars and giant planets are studied in the framework of DHOST theories. We confirm the previous results in the field that the effect of modified gravity on substellar objects' age is pronounced the most.
\end{abstract}

\maketitle

\section{Introduction}

Scalar-tensor (ST) theories of gravity are a class of modified gravity theories that were introduced in hope of finding a remedy for certain shortcomings of general relativity (GR) \cite{fuj, capo, capo2008}, such as the difficulty in explaining currently observed accelerated expansion of the Universe \cite{so2, clifton2012, so7, so3, hu2007}. ST theories are motivated not only by observational discrepancies between the theory and the empirical data, there are also some arguments coming from theories considered as fundamental, such as the string theory, which reproduce ST theory in their low-energy limit rather than GR \cite{capo2011}. 

ST theories are realized by an addition of a scalar field into the theory. Historically, the first ST theory was formulated by Brans and Dicke \cite{dicke}; their approach was later generalized to include self-interaction potential of the scalar field, yielding the so-called Wagoner parametrization \cite{jarv, wagoner}, in which the scalar field enters the action in a non-trivial way: in the most general case, it can be coupled to spacetime curvature, and to the matter fields. Such a way of introducing a new mediator of gravitational interaction results in many changes gravity manifests itself. For example, a non-minimal (anomalous) coupling between the field and matter part of the action can produce a fifth force acting on the test particles, causing them to deviate from geodesics (however, this effect can be hidden by various screening mechanisms, which are for this reason an essential part of extended gravity models \cite{screening, screening2,screening3,sunny,shei2}). On the other hand, scalar field coupled to the curvature acts as an effective gravitational constant, whose value might depend on spacetime position and matter distribution in the Universe (in agreement with Mach's principle). 

The field equations that are derived from the action considered by Brans and Dicke are of second order for the metric components and the scalar field. 
It is possible, however, to construct even more general action functional that produces second order field equations, allowing one to avoid the fatal Ostrogradsky instability which can be present when higher-order theories are considered - so-called Horndeski theory \cite{horndeski}. Furthermore, it was demonstrated that Horndeski theories are not the most general healthy ST theories: it is possible to construct higher-order theories avoiding the Ostrogradsky instability by being degenerate (so-called DHOST) \cite{langlois, langlois2}. It was discovered that the degeneracy of Lagrangian is an essential part of theories with one scalar degree of freedom allowing one to avoid the mentioned problem. 

Horndeski and DHOST theories have been widely studied in cosmology (see for a nice review and references therein \cite{kobayashi,shei1,shei3}). Interestingly, in those theories the screening mechanism turns out to be partially broken in the case of astrophysical objects \cite{broken}, such that the weak field equations are modified by the presence of terms with a numerical parameter, which encodes the information about the scalar field. This allows thus to study deviations from the Newtonian model of gravity, which is widely used to describe stellar and substellar objects. Similarly like other modified theories of gravity, the Poisson and hydrostatic equilibrium equations in such theories acquire additional terms (see \cite{Saito:2015fza,olekinv,olmo_ricci}; for review, \cite{review,cantata,anetarev}), allowing to test such theories \cite{saltas1,saltas}. Because of that, there have been a few studies undertaken which demonstrated that the Chandrasekhar mass for white dwarf stars \cite{Chandra,Saltas:2018mxc,Jain:2015edg,Banerjee:2017uwz,Wojnar:2020wd,Belfaqih:2021jvu,kalita,kalita2} may differ in modified gravity; the same happens with the minimum masses for hydrogen and deuterium burning \cite{sak1,sak2,Crisostomi:2019yfo,gonzalo,rosyadi}, as well as Jeans and opacity masses \cite{capJeans,anetaJup}. Apart from those, the evolutionary pictures of stars and planets are also distinct with respect to Newtonian gravity \cite{aneta2, chow, merce, straight,maria,debora,anetaJup2}. Those theories also modify the light elements' abundances in atmospheres \cite{aneta3} and internal structures of terrestrial planets \cite{olek,olek2,olek3}.

In the following work we are going to re-examine simple cooling models of brown dwarfs and jovian planets in the framework of {DHOST theories of gravity. It was demonstrated that both classes of these substellar objects can be a remarkable laboratory to test theoretical models of exotic particles as well as modified gravity \cite{smirnov}\footnote{Let us also notice that dark matter can be also modeled as (pseudo-)scalar field \cite{croon}.}. {Although brown dwarfs were discovered barely 30 years ago (but theorized already in 1960s), during the last years we have experienced rapidly increasing number of their discoveries, therefore an expected number of the giant exoplanets and brown dwarfs is estimated to be in billions. This provides a huge advantage with respect to neutron stars - they are easier to found and we understand physics of the processes inside of those objects much better since we deal with low temperatures and much lower densities. As it will be clearer later on, we should pay a special attention to the old objects as the electron degeneracy evolution makes that the modified gravity effects accumulate over time, overcoming the matter description uncertainties. Moreover, there are no effects resulting from the nuclear fusions as it happens in active stars. By this, one deals with additional reducing uncertainties which are caused by energy generation rates, being very sensitive to any changes in stellar cores' conditions, and by heat transport processes.} As it will be demonstrated, DHOST theories introduce extra terms to the structural equations, which can be interpreted as an additional heating or cooling process which can be not only detected, but also could explain the internal heat of the Solar System jovian planets which have not been understood well yet \cite{jdm1,jdm2,ad2}. Apart for the future constraining of the models of gravity, these objects, because of better understood physics behind their internal processes, can be an indicative tool to figure out how gravity interacts with the other forces and what is a nature of such an interplay between them.  
}

The paper is organised as follows: first, we will outline the basics of spherically-symmetric objects analysis within the paradigm of DHOST theories of gravity. This part will consist of obtaining the Lane-Emden equation, which originates from the hydrostatic equilibrium equation, and discussing the way matter is introduced into the theory. In our model, we will make use of the polytropic equation of state. The main part of the paper will be dedicated to the analysis of brown dwarfs and Jovian planets. Our aim is to study the cooling processes of those objects in the considered model of gravity, where the non-relativistic hydrostatic equilibrium equation is supplemented with an extra term. In the part dedicated to brown dwarfs, we compute time evolution of luminosity and degeneracy parameter for different values of $\Upsilon$. In case of the Jupiter, we present a modified Hertzsprung-Russell (H-R) diagram for the planet, showing how the luminosity depends on temperature for various values of the DHOST parameter. We will also briefly discuss the age of Jupiter-like planets in modified gravity. We conclude the paper with a short discussion.

\section{Basic structural and matter equations for scalar-tensor gravity}\label{basic}

As promised, firstly we will provide the basic equations to describe the stellar structure in a non-relativistic case for DHOST gravity. This will familiarize the reader with the most important formalism used to characterize the substellar objects.

\subsection{Hydrostatic equilibrium equation}

In what follows, we assume that the objects analyzed in the paper are static, spherically-symmetric, and fully convective, which allows us to make further assumptions about their matter composition. Apart from this, the substellar objects are surrounded by a radiative atmosphere with a simplified opacity model. 

{The DHOST theories can be, in general, described by the Lagrangian \cite{broken}:}
\begin{equation}
{ \mathcal{L} = M_{pl}^2\sum_i \frac{\mathcal{L}_i}{\Lambda_i^{2(i-2)}} + \alpha \phi T + \frac{T^{\mu\nu}\partial_\mu\phi\partial_\nu \phi}{\mathcal{M}^4},}
\end{equation}
{where}
\begin{subequations}
    \begin{align}
        & {\mathcal{L}_2 = X,} \\
        & {\mathcal{L}_3 = X\Box\phi -\phi_\mu \phi^{\mu\nu} \phi_\nu,} \\
        \begin{split}
             &{\mathcal{L}_4 = -X[(\Box\phi)^2 - \phi_{\mu\nu}\phi^{\mu\nu}]} \\
            &{- (\phi^\mu\phi^\nu\phi_{\mu\nu}\Box\phi - \phi^\mu \phi_{\mu\nu}\phi_\rho \phi^{\rho\nu}}
        \end{split},\\
        \begin{split}
            & {\mathcal{L}_5 = -2X[(\Box\phi)^3 - 3\phi_{\mu\nu}\phi^{\mu\nu}\Box\phi+2\phi_{\mu\nu}\phi^{\nu\rho}\phi^\mu_\rho]} \\
            & {-\frac{3}{2}\Big((\Box\phi)^2\phi^\mu\phi^\nu\phi_{\mu\nu}-2\phi_\mu\phi^{\mu\nu}\phi_{\nu\rho}\phi^\rho} \\
            & {- \phi_{\mu\nu}\phi^{\mu\nu}\phi^\rho\phi^{\rho\sigma}\phi_\sigma + 2\phi_{\mu}\phi^{\mu\nu}\phi_{\nu\rho}\phi^{\rho\sigma}\phi_\sigma\Big)}
        \end{split}
    \end{align}
\end{subequations}
{with $\Lambda_i$ being the mass-scale, $T_{\mu\nu}$ is the stress-energy tensor for matter, $X = -\frac{1}{2}\partial_\mu\phi\partial^\mu\phi$, and $\phi_{\mu_1\ldots\mu_n} = \nabla_{\mu_1}\ldots\nabla_{\mu_n}\phi$. A special subclass of these theories, called G$^3$-galileon, where the terms $\mathcal{L}_2$, $\mathcal{L}_3$ vanish; is of some interest because it yields a spherically-symmetric solution stable to perturbations \cite{broken}. This subclass is described by:}
\begin{equation}
    {L = \sqrt{-g}M_{pl}^2\left(\frac{R}{2} + X + \frac{\mathcal{L}_4}{\Lambda^4}\right)}.
\end{equation}
{In the investigations of the non-relativistic limit, one introduces the metric:
\begin{equation}
    ds^2 = -[1 + 2\Phi(r,t)]dt^2 + a^2(t)[1-2\Psi(r,t)]\delta_{ij}dx^idx^j
\end{equation}
perturbed about the FLRW metric. One also assumes that the scalar field can be decomposed as $\phi(r,t) = \phi_0(t) + \delta\phi(r,t)$. Then, introducing the new parameter \cite{broken}:
\begin{equation}
    \Upsilon = \left(\frac{\dot{\phi}_0}{\Lambda}\right)^4,
\end{equation}
one can write:}
\begin{subequations}
    \begin{align}
        & {\frac{d\Phi}{dr} = \frac{G m(r)}{r^2} + \frac{\Upsilon}{4}G_N m''(r),}\label{first} \\
        & {\frac{d\Phi}{dr} = \frac{G m(r)}{r^2} -\frac{5\Upsilon}{4}\frac{G_N m'(r)}{r}.}
    \end{align}
\end{subequations}
{It is the first quantity, \eqref{first}, that is related to the motion of non-relativistic particles, and for this reason, we will focus on it. Then, we have:
\begin{equation}
    \frac{1}{\rho}\frac{dp}{dr}=-\frac{d\Phi}{dr},
\end{equation}
so} the non-relativistic hydrostatic equilibrium equation in the {DHOST theories} can be written as \cite{broken}:
\begin{equation}\label{hydro}
    \frac{dp}{dr} = -\frac{Gm(r)}{r^2}\rho(r) - \frac{\Upsilon}{4}G \rho(r)\frac{d^2m(r)}{dr^2}
\end{equation}
where $\Upsilon$ denotes a parameter characterizing the theory. In what follows, we will consider its values from the range $-\frac{2}{3}<\Upsilon\lesssim 1.4$ \cite{sak1} {and also much tighter range, provided by helioseismology: $-10^{-3}<\Upsilon\lesssim 5\times 10^{-4}$ \cite{saltas}}. {The first range was obtained from analyzing brown dwarf stars physics and they observational properties, taking into account a description of the electron degeneracy properties which we need in our studies even more detailed. To obtain the second range, the authors were using a simplified description of the Sun, although they provided a very rigorous uncertainties analysis which the first one was lacking. Because of that reason, we will focus on those two ranges, and at the end we will also perform the uncertainties analysis.}

The mass function in the considered theory has the following, well-know relation:
\begin{equation}
   \frac{dm}{dr}= 4\pi r^2\rho(r).
\end{equation}

In order to model the interior and the atmosphere of an astrophysical object, one also needs to specify the type of heat transport. Usually, one uses the Schwarzschild criterion \cite{schw,schw2} to determine whether one deals with either a diffusive/conductive transport, or with an adiabatic convection present locally. One introduces the temperature gradient:
\begin{equation}
    \nabla_{rad} = \left(\frac{d\ln T}{d\ln p}\right)_{rad},
\end{equation}
which, for our theory, can be written as \cite{debora}:
\begin{equation}
    \frac{\partial T}{\partial p}= \frac{3\kappa_\text{rc}L}{16\pi G M \bar ac T^3}\left(1+\frac{\Upsilon}{2}\right)^{-1},
\end{equation}
{where we have used the fact that the surface gravity $g$ can be treated as a constant (see the eq. \eqref{surgrav})},
such that
\begin{equation}\label{schwcrit}
    \nabla_\text{rad}=\frac{3\kappa_\text{rc}Lp}{16\pi G M \bar ac T^4}\left(1+\frac{\Upsilon}{2}\right)^{-1}.
\end{equation}
The constant $\bar a = 7.57\times 10^{-15}\frac{erg}{cm^3K^4}$, the radiative and/or conductive opacity is denoted by $\kappa_\text{rc}$, and $L$ is the luminosity. In order to determine the kind of the heat transport we are dealing with,
\begin{align*}
 \nabla_\text{rad}\leq&\nabla_{ad}\;\;\text{\small pure diffusive radiative or conductive transport}\\
  \nabla_\text{rad}>&\nabla_{ad}\;\;\text{\small adiabatic convection is present locally}.
\end{align*}
one compares it with the adiabatic gradient, $\nabla_{ad}$, whose value depends on the properties of the gas. For an ideal gas, the adiabatic gradient becomes $\nabla_{ad} =0.4$. 

In order to solve the set of differential equations, we need to impose an additional relation between the pressure and the energy density. We will work with (analytical) barotropic equations of state
\begin{equation}
    p=p(\rho),
\end{equation}
whose polytropic form $p=K\rho^\frac{n+1}{n}$ will be discussed in more detail in the upcoming subsection.
For the polytropic equation of state, it is convenient to use the Lane-Emden (LE) approach, allowing one to write the equations with dimensionless quantities $\theta$ and $\xi$. For the {DHOST} theories, the LE equation reads as \cite{broken,sak1,sak2}: 
\begin{equation}\label{LE}
    \frac{1}{\xi^{2}} \frac{\mathrm{d}}{\mathrm{d} \xi}\left[\left(1+\frac{n}{4} \Upsilon \xi^{2} \theta^{n-1}\right) \xi^{2} \frac{\mathrm{d} \theta}{\mathrm{d} \xi}+\frac{\Upsilon}{2} \xi^{3} \theta^{n}\right]=-\theta^{n},
\end{equation}
with the following definitions:
\begin{equation}\label{radius_density}
    r = r_c \xi, \quad \rho = \rho_c \theta^n, \quad p = p_c\theta^{n+1}, \quad r_c^2=\frac{(n+1)p_c}{4\pi G\rho_c^2},
\end{equation}
where $p_c$ and $\rho_c$ are the values of pressure and density at the core of the object, respectively. One can solve the equation numerically and get the star's radius, mass, temperature and central density:
\begin{subequations}
\begin{align}
    &R = \gamma_n \left(\frac{K}{G}\right)^{\frac{n}{3-n}}M^\frac{n-1}{n-3}, \quad M = 4\pi r_c^3 \rho_c \omega_n, \label{radius0}\\
    & T = \frac{K\mu}{k_B}\rho_c^{\frac{1}{n}}\theta_c, \quad \rho_c = \delta_n \frac{3M}{4\pi R^3},
\end{align}
\end{subequations}
where $\mu$ denotes the mean molecular mass and $k_B$ is the Boltzmann constant. The parameters $\{\omega_n, \gamma_n, \delta_n\}$ depend on the theory of gravity and are defined as follows:
\begin{align}
\omega_{n}  &=-\left.\xi^{2} \frac{d \theta}{d \xi}\right|_{\xi=\xi_{R}},\,\,\,
\delta_{n}  =-\frac{\xi_{R}}{3 d \theta /\left.d \xi\right|_{\xi=\xi_{R}}}\\
\gamma_{n}  &=(4 \pi)^{\frac{1}{n-3}}(n+1)^{\frac{n}{3-n}} \omega_{n}^{\frac{n-1}{3-n}} \xi_{R},
\end{align}
where $\theta(\xi)$ is a solution of \eqref{LE}, while $\xi_R$ is a value for which $\theta(\xi_R)=0$, that is, the first zero indicates the radius of the object. { In the further part, we will skip the index $n$ in the Lane-Emden parameters $\omega$, $\gamma$, and $\delta$.}

The luminosity of a substellar object is given by the Stefan-Boltzmann law
\begin{equation}\label{stefan}
    L=4\pi R^2\sigma T^4_{eff},
\end{equation}
where $R$ is given by the solution of the structural equation given by \eqref{radius0} while the effective temperature $T_\text{eff}$, in our assumption being also the photosphere temperature, must be determined by other meanings, discussed briefly in the brown dwarfs' section and derived in the jovian planets' one.

\subsection{Matter description}\label{matter}

To describe matter our substellar object is made of, we make use of the polytropic equation of state given in the following form:
\begin{equation}\label{poly}
    p = K\rho^{1 + \frac{1}{n}},
\end{equation}
where $n$ is a constant, polytropic index, whose value depends on a type of an objects while $K$ provides the information on the composition of matter and its properties, for instance, interactions between particles and electron degeneracy. In the simplest case, $K$ is a constant depending on $n$, such that in our case it is given for $n=\frac{3}{2}$ by\footnote{The polytropic parameter $n=3/2$ describes a non-relativistic electron gas which is good enough to model fully convective objects.}
\begin{equation}
    K = \frac{1}{20}\left(\frac{3}{\pi}\right)^{\frac{2}{3}}\frac{h^2}{m_e}\frac{1}{(\mu_e m_u)^\frac{5}{3}},
\end{equation}
where $m_e$ is the electron mass, $\mu_e$ is related to the number of baryons per electron: $\frac{1}{\mu_e} = X + Y/2$, with X and Y being
the mass fractions of hydrogen and helium, respectively, and $m_u$ is the mass of a nucleon. We will use a mixture of this simplified equation of state with ideal gas in order to model a Jupiter-like planet.

On the other hand, when brown dwarf stars considered, one needs to take into account the additional effects arising 
when a mixture of the degenerate Fermi gas of electrons at a finite temperature with a gas of ionized hydrogen and helium is considered. It turns out that the resulting equation of state can be also written in the form of a polytrope with the polytropic index $n=3/2$ \cite{aud}, however the polytropic parameter $K$ takes the form
\begin{equation}\label{ka}
K=C \mu_{e}^{-5 / 3}(1+b+ a \Psi),
\end{equation}
where
\begin{align}
b&=-\frac{5}{16} \Psi \ln \left(1+e^{-\frac{1}{\Psi}}\right)+\frac{15}{8} \Psi^{2}\left\{\frac{\pi^{2}}{3}+\mathrm{Li}_{2}\left(-e^{-\frac{1}{\Psi}}\right)\right\}\\
 a&=\frac{5\mu_e}{2\mu_1},\;\;\;\;\frac{1}{\mu_1}=(1+x_{H^+})X+\frac{Y}{4} ,
\end{align}
where
$\mu_1$ is the mean molar mass for ionised hydrogen and helium mixture with $x_{H^+}$ being the ionization fraction of hydrogen, $Li_2(x)$ is a polylogarithmic function, while
the constant $C = 10^{13}$ cm$^4$ g$^{-2/3}$ s$^{-2}$. The quantity $\Psi$ is the degeneracy parameter, which is defined as
\begin{equation}\label{degeneracy}
\Psi=\frac{k_{B} T}{\mu_{F}}=\frac{2 m_{e} k_{B} T}{\left(3 \pi^{2} \hbar^{3}\right)^{2 / 3}}\left[\frac{\mu_{e}}{\rho N_{A}}\right]^{2 / 3},
\end{equation}
where $N_A$ is the Avogadro number while the other constants have the standard meaning.

Another thermodynamic quantity is the internal entropy. It was showed that in the case of brown dwarfs with the interior described by the above equation of state is given by
\begin{equation}\label{entropy}
 S_\text{interior}=\frac{3}{2}\frac{k_BN_A}{\mu_{1mod}}(\rm{ln}\Psi+12.7065)+C_1,
\end{equation}
where $C_1$ is an integration constant of the first law of thermodynamics and
\begin{equation}
 \frac{1}{\mu_{1mod}}=\frac{1}{\mu_1}+\frac{3}{2}\frac{x_{H^+}(1-x_{H^+})}{2-x_{H^+}},
 \end{equation}
with $x_{H^+}$ being the ionization fraction of hydrogen. In the further part, we will consider a specified metallic-molecular phase transition model with $x_{H^+}=0.255$, given in \cite{aud}.

When analyzing the atmosphere of substellar objects, it is convenient to introduce the so-called optical depth:
\begin{equation}\label{depth}
    \tau(r) = \int^\infty_r \bar{\kappa}\rho dr,
\end{equation}
where $\bar{\kappa}$ denotes the mean opacity. To describe objects whose surface temperature is low, one can use Rosseland mean opacities given by Kramer's law:
\begin{equation}\label{kram}
    \bar{\kappa} = \kappa_0 p^u T^w
\end{equation}
with $\kappa_0$, $u$ and $w$ being constants whose values depend on the opacity regime. Regarding the composition of the atmosphere, we assume that it can be modelled as ideal gas with the equation of state:
\begin{equation}\label{ideal}
    \rho = \frac{\mu p}{N_A k_B T}.
\end{equation}

\section{Brown dwarf stars}
Before going to the theoretical description of the brown dwarf stars, let us briefly recall the basic notions related to those failed stars. According to the current models based on Newtonian gravity, they are objects with masses from the range $(\sim 0.08-\sim0.003M_\odot)$ \cite{cab3,boss}; the upper limit corresponds to the minimum mass for hydrogen burning\footnote{That is, roughly speaking, the mass an object needs to have in order to star hydrogen fusion in its core which results as a counterbalance process to the gravitational contraction. Such a star enters then the Main Sequence phase.} \cite{burrows1,burrows2} while the lower one is related to the so-called opacity mass \cite{rees}. The opacity mass limit is the smallest mass bounded gravitationally which cools via radiation processes. Using other words, it is the smallest mass of a gas cloud which will not crumble into smaller pieces caused only by gravitational instabilities. It is believed that stars and brown dwarfs form via such a fragmentation process, which is limited by the opacity mass, while gaseous giant planets are made from the gas and rocks of the protoplanetary disc surrounding just a formed parent star.

Let us notice that the above boundary masses as well as mentioned processes related to the substellar formation depend on the interior structure, first-order phase transition, opacity and atmosphere model \cite{aud}, as well as modified gravity \cite{sak1,sak2,Crisostomi:2019yfo,gonzalo,rosyadi,capJeans,merce,maria,debora,anetaJup,anetaJup2}. Moreover, massive brown dwarfs can burn deuterium as well as lithium and even hydrogen in their cores, however those processes are neither stable nor energetic enough to stop the gravitational contraction. Therefore, they continue shrinking, radiating the stored energy away and cooling down with time. The contraction stops on the onset of the electron degeneracy.

In what follows, we will model our brown dwarf star as a ball with two layers: the interior which is described by the equation of state \eqref{poly} with $K$ given by \eqref{ka}, and atmosphere whose matter properties are given by the ideal gas relation \eqref{ideal} and opacity \eqref{kram}.

\subsection{Theoretical framework}
With the use of the Lane-Emden formalism presented in the section \ref{basic} we may write down the radius', central density's and pressure's dependence on the brown dwarf's mass and electron degeneracy:
\begin{equation}\label{radius}
    R=1.19138\times10^9\gamma\left(\frac{M_\odot}{M}\right)^{1/3}\mu_e^{-5/3}(a\Psi + b + 1)[cm]
\end{equation}
\begin{equation}\label{central density}
    \rho_c=2.808007\times10^5\frac{\delta}{\gamma^3}\left(\frac{M}{M_\odot}\right)^{2}\frac{\mu_e^{5}}{(a\Psi + b + 1)^3}[g/cm^3]
\end{equation}
\begin{equation}\label{central_pressure}
    p_c=1.204103\times10^{10}\frac{\delta^{5/3}}{\gamma^5}\left(\frac{M}{M_\odot}\right)^{10/3}\frac{\mu_e^{20/3}}{(a\Psi + b + 1)^4}[Mbar],
\end{equation}
while the central temperature is given by combining the equations \eqref{degeneracy} and \eqref{central density}:
\begin{equation}\label{central_temp}
    T_c=1.294057\times10^{9}\frac{\delta^{2/3}}{\gamma^2}\left(\frac{M}{M_\odot}\right)^{4/3}\frac{\Psi\mu_e^{8/3}}{(a\Psi + b + 1)^2}[K].
\end{equation}
On the other hand, modelling the surface properties of those objects requires knowledge on a first order phase transition between the interior, characterized by a mixture of metallic hydrogen and helium, and photopshere, with molecular hydrogen and helium composition. Following the result given by \cite{aud,chab,chab2}, the effective temperature can be expressed as
\begin{equation}\label{teff}
    T_{eff}=b_1\times10^6\rho_{e}^{0.4}\psi^\nu\ [\text{K}],
\end{equation}
where $b_1$ and $\nu$ are numerical values\footnote{Their values, together with the values for the ionization fraction of hydrogen $x_{H^+}$ can be found in \cite{aud,maria}. { In the further part, as an example, we will focus on the model D, so $b_1=2$ and $\nu = 1.6$, while $x_{H^+}=0.255$.}} depending on the phase transition.

In order to follow further, we assume that the photosphere's radius is approximately equalled to the radius of the brown dwarf star; moreover, the surface gravity $g$ can also be taken as a constant value,
\begin{equation}\label{surgrav}
    g=\frac{Gm(r)}{r^2}=\text{const}
\end{equation}
such that the hydrostatic equilibrium equation \eqref{hydro} at the photosphere can be written as \cite{sak1,sak2}
\begin{equation}\label{hyd2}
    p'=-g\rho\left(1+\frac{\Upsilon}{2}\right).
\end{equation}
The photosphere is defined at the radius when the optical depth given by \eqref{depth} is equalled $2/3$
\begin{equation}
    \tau(r)=\kappa_R\int_r^\infty \rho dr=\frac{2}{3}.
\end{equation}
with the Rosseland's mean opacity $\kappa_R=0.01\text{cm}^2/\text{g}$. Using this definition and the photopsheric hydrostatic equilibrium given by \eqref{hyd2}, one can find the photospheric pressure
\begin{equation}
    p_{ph}=\frac{2GM}{3\kappa_R R^2}\left(1+\frac{\Upsilon}{2}\right).
\end{equation}
Inserting the radius relation \eqref{radius} into above yields (in [bar])
\begin{equation}\label{fotpres}
    p_{ph}=\frac{62.352023}{\kappa_R\gamma^2}\left(\frac{M}{M_\odot}\right)^{5/3}\frac{\mu_e^{10/3}}{(a\Psi+b+1)^2}\left(1+\frac{\Upsilon}{2}\right).
\end{equation}
Since in our model the photosphere's equation of state is given by the ideal gas \eqref{ideal}, the photospheric density can be easily obtained and inserted in \eqref{teff}, such that the effective temperature is now written as
\begin{align}\label{tefflum}
\begin{split}
    T_{eff}=& \frac{2.557879\times10^4}{\kappa_R^{2/7}\gamma^{4/7}}\left(\frac{M}{M_\odot}\right)^{10/21}\\
  &\times\frac{b_1^{5/7}\Psi^{\nu\cdot5/7}}{(a\Psi+b+1)^{4/7}}\left(1+\frac{\Upsilon}{2}\right)^{2/7}[K].
\end{split}
\end{align}
Finally, the luminosity of the brown dwarf stars is obtained as a function of its mass and electron degeneracy $\Psi$ by inserting the above effective temperature together with \eqref{radius}
\begin{align}\label{lum}
    L&=\frac{0.072233L_\odot}{\kappa_R^{8/7}\gamma^{2/7}}\left(\frac{M}{M_\odot}\right)^{26/21}
  \frac{b_1^{20/7}\Psi^{\nu\cdot20/7}}{(a\Psi+b+1)^{2/7}}\left(1+\frac{\Upsilon}{2}\right)^{8/7}.
\end{align}
This is the main result related to the modelling of brown dwarf stars in DHOST, where the modification with respect to the Newtonian model is given by the presence of the parameters $\Upsilon$ and $\gamma$. Let us however notice that those objects undergo the gravitational contraction, since there is no energy source whose pressure could counterbalance the attraction, apart from the initial unstable hydrogen burning in the case of the very massive brown dwarfs. Depletion of lithium and deuterium in massive brown dwarfs is not sufficient to stop the contraction and in our approximation the energy generated by these nuclear processes can be ignored, therefore those bodies will cool down with time. The electron degeneracy is however the non-negligible effect in the cooling process, and its evolution while a brown dwarf contracts should be also taken into account.

In order to find the time dependency of the electron degeneracy $\Psi$, let us consider the energy equation, given by
\begin{equation}
    \frac{dE}{dt}+p\frac{dV}{dt}=T\frac{dS}{dt}=\dot{\epsilon}-\frac{\partial L}{\partial M},
\end{equation}
where $E$ is the energy of the system, $V$ the volume, $S$ the entropy per unit mass, and $L$ the surface luminosity. As mentioned, we may neglect the energy generation term $\dot{\epsilon}$, such that integrating over mass the last two terms from the above equation one finds
\begin{equation}
\frac{{d} s}{\mathrm{dt}}\left[\int N_{A} k_{B} T {dM}\right]=-L
\end{equation}
where we have defined $s=S/k_B N_A$. The polytropic equation of state \eqref{poly} inserted to \eqref{degeneracy} allows to get rid of the temperature from the previous expression, such that
\begin{equation}\label{sigma}
    \frac{{d} s}{{dt}} \frac{N_{A} A \mu_{e} \Psi}{C(1+b+a \Psi)} \int p {dV}=-{L},
\end{equation}
where the constant was defined $A=\frac{(3N_A\pi^2\hbar^3)^{2/3}}{2m_e}$ while the integral is given by
\begin{equation}\label{vol}
    \int p {d} {V}=\frac{2}{7} {G} \frac{{M}^{2}}{{R}}.
\end{equation}
The entropy rate can be obtained from the relation \eqref{entropy}
\begin{equation}
    \frac{{d} s}{{dt}}=\frac{1.5}{\mu_{1 \bmod }} \frac{1}{\Psi} \frac{{d} \Psi}{{dt}},
\end{equation}
which inserted together with \eqref{vol} and luminosity \eqref{lum} to \eqref{sigma} provides the degeneracy's evolution 
\begin{align}\label{deg_ev}
\begin{split}
   \frac{d\Psi}{dt}&=\frac{-1.018097\times10^{-18}\mu_{1mod}}{\kappa_R^{8/7}}\left(\frac{M_\odot}{M}\right)^{23/21}\\
&\times\frac{b_1^{20/7}\Psi^{\nu\cdot20/7}(a\Psi+b+1)^{12/7}}{\mu_e^{8/3}}\left(1+\frac{\Upsilon}{2}\right)^{8/7}, 
\end{split}
\end{align}
which clearly also depends on the theory parameter.

\subsection{Numerical solutions}
\begin{figure}[t]
\centering
\includegraphics[scale=0.8]{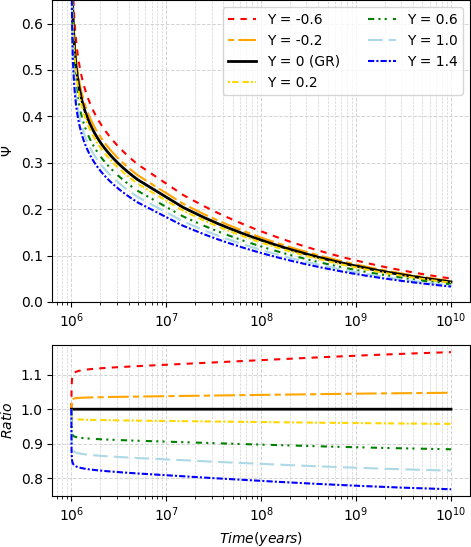}
\caption{[color online] The time evolution of the degeneracy parameter $\Psi$ for $M={0.05}{M_\odot}$ and different values of the $\Upsilon$ parameter. The bottom panel shows the ratio of the time evolution in the scalar-tensor gravity with respect to $\Upsilon=0$.}\label{deg1}
\end{figure}

\begin{figure}[t]
\centering
\includegraphics[scale=0.65]{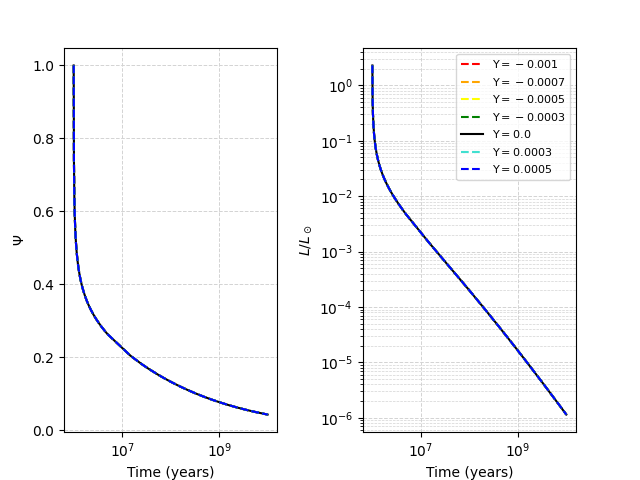}
\caption{{[color online] The time evolution of the degeneracy parameter $\Psi$ and the luminosity (as compared to Sun's luminosity) for $M={0.05}{M_\odot}$ and different values of the $\Upsilon$ parameter. Clearly, the effect of modified gravity is negligible for such a range of the parameter $\Upsilon$.}}\label{deg2}
\end{figure}

\begin{figure}[t]
\centering
\includegraphics[scale=0.9]{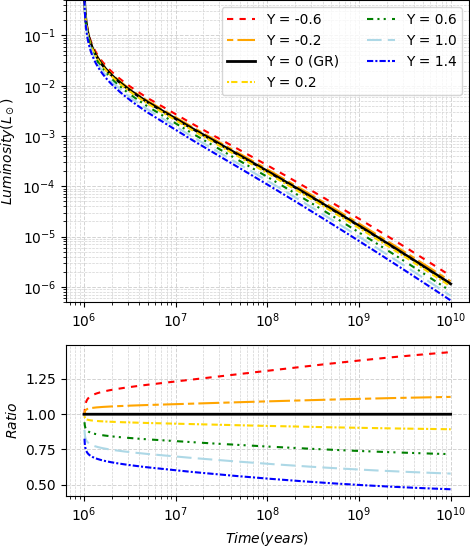}
\caption{[color online] The time evolution of a $M={0.05}{M_\odot}$ brown dwarf's luminosity.  The bottom panel shows the ratio of the time evolution in the scalar-tensor gravity with respect to $\Upsilon=0$.}
\label{tlum}
\end{figure}

The equation \eqref{deg_ev} is numerically solved with the initial condition $\Psi=1$ at $t=0$ for the theory parameter $\Upsilon$ from the range $[-0.6; 1.4]$ {(see Fig \ref{deg1}}), where $\Upsilon = 0$ provides the Newtonian model, {and for a much tighter range $[-10^{-3};5\times 10^{-4}]$ (see Fig \ref{deg2}), as suggested by \cite{saltas}}. { The choice of the initial value of the degeneracy parameter comes from the assumption that, at the beginning, when the object is large, there is no degeneracy (one needs to remember that smaller values of the parameter $\psi$ correspond to a bigger extent of degeneracy, see the definition \eqref{degeneracy}). When integrating the equations, we assumed that $u = 0.5$ and $w = 1$ in \eqref{kram}, as suggested by the work \cite{don}. The value of $\kappa_0 = 10^{75}$ in \eqref{kram} was selected in such a way that the model reproduced cooling time of the Jupiter in the case of Newtonian gravity, that is, when $\Upsilon = 0$. The mass fraction of hydrogen and helium are set to be $X = 0.75$ and $Y = 0.25$, respectively, so $\mu_e=1.14$.} The solutions are given by the figures \ref{deg1} and \ref{deg2}. Using these results in \eqref{lum}, one finally gets the luminosity as a function of time, given by the figure \ref{tlum}.

\section{Jovian planets}\label{jovian}
There is no doubt that jovian planets possess a complicated internal structure, as indicated by by theoretical models and Juno mission collecting data on Jupiter \cite{jeff,jup,jup1,jup2,jup3,jup4,jup5,juno,juno2,juno3,stev_juno}. It seems that the internal model which goes well with the current observational data must have at least 3 layers: a diffusive core built of heavier elements, a mantle mainly composed of metallic hydrogen with admixtures of helium and heavier elements. The outer layer mainly consists of molecular hydrogen with helium rain and silicate droplets \cite{helled}. Apart from such a complex structure, there is still no consistent equation of state describing a mixture of hydrogen and helium in the pressure range approaching megabar, while the temperature can reach even a few thousand kelvins \cite{stev_juno,mix1,mix2}. Despite this, we can still model a Jupiter-like planet with the use of a simplified and analytical equation of state, as presented further, because it was demonstrated that it can produce a cooling model with the nowadays' surface temperature close enough to the actual value \cite{don,don0}. Since the model is simple, it also allows us to trace differences provided by the scalar-tensor gravity with respect to the Newtonian one. Therefore, we will follow the derivation of the atmospheric, boundary, and convective interior given by \cite{don} and further studied by \cite{anetaJup}.

\subsection{Atmosphere quantities for the jovian planets}

The planet's luminosity is described by the Stefan-Boltzmann law. In general, there can be various energy sources contributing to the total radiation energy of the planet. To simplify our considerations, we assume here that the only relevant energy source of energy for a given planet is the energy flux from the nearest star:
\begin{equation}
    L_\text{received}=\left(\frac{R_\text{p}}{2R_\text{sp}}\right)^2L_\text{s},
\end{equation}
where $R_p$ denotes planet's radius, $L_s$ luminosity of the star, and $R_{sp}$ is the distance between the planet and the star. Some part of the energy is directly reflected by the planet, and the absorbed energy is given by the formula:
\begin{equation}\label{abs}
    L_\text{abs}=(1-A_\text{p})\left(\frac{R_\text{p}}{2R_\text{sp}}\right)^2L_\text{s}.
\end{equation}
where $A_p$ is the plant's albedo. Making the assumption that the distribution of the energy absorbed is uniform, we can obtain the equilibrium temperature $T_{eq}$ by using the Stefan-Boltzmann law:
\begin{equation}
        (1-A_\text{p})\left(\frac{R_\text{p}}{2R_\text{sp}}\right)^2L_\text{s}=4\pi f\sigma T_\text{eq}^4R_\text{p}^2,
\end{equation}
with $f<1$ being a factor which allows to take into account the fact that the planet radiates less energy than the black body with the same effective temperature. 
The thermal equilibrium is achieved when the energy received from the star becomes equal to the energy planet radiates away from its surface. This allows one to write $T_{eff} = T_{eq}$. Taking into account the fact that the star's luminosity is given by:
\begin{equation}
    L_\text{s}=4\pi\sigma T_\text{s}^4R_\text{s}^2
\end{equation}
we are able to relate the equilibrium temperature to the star's surface temperature $T_s$:
\begin{equation}\label{teq}
  T_\text{eq}=  (1-A_\text{p})^\frac{1}{4}\left(\frac{R_\text{s}}{2R_\text{sp}}\right)\frac{1}{2}T_\text{s}.
\end{equation}
Interestingly, $T_{eq}$ does not depend on the planet's size in this case. These considerations are not true for more general cases when one needs to take into account internal sources of energy, such as gravitational contraction, tidal forces, or Ohmic heating. With these additional sources, the planet's temperature is higher than $T_{eq}$, and the planet radiates away more energy than it receives from the star. To find the relation between the effective and equilibrium temperatures, one can use the standard equation describing radiative transfer in grey atmosphere \cite{saumon,guillot, seager} and Eddington's approximation. One can show that \cite{don}: 
\begin{equation}\label{temp}
 4T^4=3\tau(T^4_\text{eff}-T^4_\text{eq})+2(T^4_\text{eff}+T^4_\text{eq}).
\end{equation}
Here, $T$ denotes the stratification temperature in the atmosphere, {described  by
\begin{equation}
    \frac{dT}{dr} = \frac{3\bar{\kappa} L \rho}{64\pi\sigma r^2 T^3}
\end{equation} }and $\tau$ is the optical depth. Zero value of the depth is achieved at the planet's surface. The equation can be rewritten in a simpler form if one introduced the following quantities:
\begin{equation*}
    T_-:=T^4_\text{eff}-T^4_\text{eq},\;\;\;T_+:=T^4_\text{eff}+T^4_\text{eq}
\end{equation*}
so that the equation (\ref{temp}) reads now:
\begin{equation}\label{temp2}
 4T^4=3\tau T_-+2T_+.
\end{equation}
We can use the fact that the atmosphere is in hydrostatic equilibrium with gravitational pressure to find the pressure in the atmosphere. The optical depth definition can be used in the hydrostatic equilibrium equation to relate the pressure to the density and gravitational interaction:
\begin{equation}\label{equil}
 \frac{dp}{dr}=-\kappa\rho\frac{dp}{d\tau}=-g\rho\left(1+\frac{\Upsilon}{2}\right).
\end{equation}
The opacity is given by the equation \eqref{kram} with $u$ and $w$ unspecified to make our considerations more general. Now, using \eqref{kram}, we can rewrite \eqref{equil} as:
\begin{equation}\label{prestemp}
 p^u \frac{dp}{d\tau}=\frac{g}{\kappa_0 T^{4w}}\left(1+\frac{\Upsilon}{2}\right).
\end{equation}
which can be plugged in (\ref{temp}) to give:
\begin{equation}
 \int^p_0 p^u dp=\frac{4^w g}{\kappa_0}\left(1+\frac{\Upsilon}{2}\right)\int^\tau_0\frac{d\tau}{(3\tau T_-+2T_+)^w}.
\end{equation}
We integrate it for $w\neq 1$ and $w=1$, respectively, to get the pressure in the atmosphere:
\begin{align}\label{presat}
\begin{split}
     p^{u+1}=&\frac{4^wg}{3\kappa_0}\frac{u+1}{1-w}\left(1+\frac{\Upsilon}{2}\right)
 T_-^{-1}\Big((3\tau T_-+2T_+)^{1-w} \\
 & -(2T_+)^{1-w}\Big),
\end{split}\\
 p^{u+1}=&\frac{4g}{3\kappa_0}(u+1)\left(1+\frac{\Upsilon}{2}\right)T_-^{-1}\text{ln}[3\tau T_-+2T_+],
\end{align}
where we have used the boundary condition $p=0$ at $\tau=0$.

\subsection{Boundary between the radiative atmosphere and convective interior}

Inside the gaseous planets, the transport of energy can be attributed to convective processes. Between the atmosphere and the interior of the planet, the transport of energy is replaced by the radiative one. The change of the type of energy transport can be described by the Schwarzschild criterion \eqref{schwcrit}. The behavior of the convective interior can be modelled with a polytropic equation of state \eqref{poly} with $n=3/2$, so that the stratification $d\ln{T}/d\ln{p}=\nabla_\text{ad}$ is adiabatic and equal to $2/5$ for fully ionised gas \cite{stellar}. Using \eqref{prestemp} together with the Schwarzschild condition, we get:
\begin{equation}
    \frac{15}{32}p^{u+1}T^{-4}T_-=\frac{g}{\kappa_0 T^{4w}}\left(1+\frac{\Upsilon}{2}\right),
\end{equation}
which, upon substituting the temperature of the atmosphere \eqref{temp2} and its pressure \eqref{presat}, gives the critical depth:
\begin{align}
 \tau_c&=\frac{2}{3}\frac{T_+}{T_-}\left(\Big(1+\frac{8}{5}\Big(\frac{w-1}{u+1}\Big)\Big)^\frac{1}{w-1}-1\right),\;\;w\neq1\\
 \tau_c&=\frac{2}{3}\frac{T_+}{T_-}(e^\frac{16}{15}-1),\;\;w=1.
\end{align}
At this optical depth, the radiative transport is replaced by the convective one. To find the temperature and pressure at the radiative-convective boundary, one substitutes the relations given above to \eqref{temp2} and \eqref{presat}:

\begin{align}\label{pressure_boundaryy}
 p^{u+1}_\text{conv}=&\frac{8g}{15\kappa_0}\frac{4^w\left(1+\frac{\Upsilon}{2}\right)}{T_-(2T_+)^{w-1}}\left(\frac{5(u+1)}{5u+8w-3}\right),\\
 T^4_\text{conv}=&\frac{T_+}{2} \left(\frac{5u+8w-3}{5(u+1)}\right)^{w-1}\label{t1}
\end{align}

for $w\neq1$. For $w=1$, those equations give:
\begin{align}
 p^{u+1}_\text{conv}=&\frac{32g}{15\kappa_0}\frac{\left(1+\frac{\Upsilon}{2}\right)}{T_-},\\
 T^4_\text{conv}=&\frac{1}{2}T_+e^\frac{16}{15}.\label{t2}
\end{align}

\subsection{Convective interior of the Jovian planets}

In the following parts of the paper, when modelling the interior pressure of Jovian planets, we will assume it can be split into two parts:
\begin{equation}\label{prescomb}
    p=p_1+p_2,
\end{equation}
where $p_1$ comes from the electron degeneracy and is given by the polytropic EoS \eqref{poly} with $n=3/2$, and $p_2$ is simply the pressure of an ideal gas:
\begin{equation}
    p_2=\frac{k_B\rho T}{\mu},
\end{equation}
where $\mu$ denotes the mean molecular weight. It can be shown that such a combination of pressures can be described by a single polytropic EoS:
\begin{equation}
    p = A\rho^\frac{5}{3},
\end{equation}
where $A = p_c / \rho_c^\frac{5}{3}$ (index denotes values at the core of the object). We can then substitute it in \eqref{prescomb} and make use of the Lane-Emden relations \eqref{central density} and \eqref{central_pressure} to obtain:
\begin{equation}
 A=\gamma^{-1}GM_p^\frac{1}{3}R_p.
\end{equation}
The interior pressure \eqref{prescomb} can be now written as:
\begin{equation}\label{p2}
 p_\text{conv}=\frac{GM_p^{1/3}R_p}{\gamma}\left(\frac{kT_\text{conv}}{\mu\left(G\gamma^{-1}M_p^{1/3}R_p-K\right)}\right)^\frac{5}{2}.
\end{equation}
This pressure must be equal to the pressure at the radiative-convective boundary \eqref{pressure_boundaryy}:
\begin{align}\label{cond}
\begin{split}
    & T_+^{\frac{5}{8}u+w-\frac{3}{8}}T_-=\bar CG^{-u} M_p^{\frac{1}{3}(2-u)}R_p^{-(u+3)}\mu^{\frac{5}{2}(u+1)}k_B^{-\frac{5}{2}(u+1)}
\gamma^{u+1}\\
&\times(G\gamma^{-1}M_p^\frac{1}{3}R_p-K)^{\frac{5}{2}(u+1)}\left(1+\frac{\Upsilon}{2}\right).
\end{split}
\end{align}
Using this condition, one can relate the effective temperature $T_\text{eff}$ to the radius of the planet $R_p$. In the equation above, $\bar C$ is a constant whose value depends on the opacity constants $u$ and $w$, $w>1$:

\begin{align}
 \bar C_{w\neq1}=&\frac{16}{15\kappa_0}2^{\frac{5}{8}(1+u)+w}\left(\frac{5u+8w-3}{5(u+1)}\right)^{1+\frac{5}{8}(1+u)(w-1)},\\
 \bar C_{w=1}=&\frac{32}{15\kappa_0}2^{\frac{5}{8}(u+1)}e^{-\frac{2}{3}(u+1)}.
\end{align}

The final radius of the planet can be obtained from the Eq. \eqref{cond} after setting $T_- = 0$; this condition means that the effective planet's temperature reached an equilibrium value so that the only source of energy for the fully contracted planet is the parent star. The radius of a contracted planet is given by:

\begin{equation}
 R_F=\frac{K\gamma}{GM_p^\frac{1}{3}}.
\end{equation}

The effect of the non-relativistic limit of { DHOST theories} is contained within the $\gamma$ parameter. For different values of the theory parameter $\Upsilon$, the final radius will be either larger or smaller than the one predicted by Newtonian gravity. 

\subsection{Jovian planets' evolution}

We assume that the process of contraction is quasi-equilibrium, which allows us to write the planet's luminosity as a sum of the internal gravitational energy and the total energy absorbed by the planet, $L_\text{abs}$. For a polytrope with the polytropic index $n=3/2$, we can write \cite{maria}:

\begin{equation}\label{cooling}
 L_p=L_\text{abs}-\frac{3}{7}\frac{GM_p^2}{R_p^2}\frac{dR_p}{dt}.
\end{equation}

Using the Stefan-Boltzmann law \eqref{stefan} and Eq. \eqref{abs}, the evolution equation given above can be written as:

\begin{equation}
 \pi \bar a c R^2_pT_-=-\frac{3}{7}\frac{GM_p^2}{R_p^2}\frac{dR_p}{dt}.
\end{equation}

In order to obtain the contraction time, one needs to integrate this equation from the initial radius $R_0$ to the final one $R_F$:

\begin{equation}
 t=-\frac{3}{7}\frac{GM_p^2}{\pi \bar ac}\int^{R_p}_{R_0}\frac{dR_p}{R_p^4T_-}.
\end{equation}

Here, $T_-$ can be thought of as a function of $R_p$ and $T_\text{eff}$ (cf. \eqref{cond}). Thus, we can write the integral in the following way:

\begin{align}\label{age}
\begin{split}
    t=& -\frac{3}{7}\frac{GM_p^\frac{4}{3}k_B^{\frac{5}{2}(u+1)}\kappa_0}{\pi \bar ac\gamma\mu^{\frac{5}{2}(u+1)}K^{\frac{3}{2}u+\frac{5}{2}}\bar C}\left(1+\frac{\Upsilon}{2}\right)^{-1}\\
&\times\int^{1}_{x_0}\frac{(T_\text{eff}^4+T^4_\text{eq})^{\frac{5}{8}u+w-\frac{3}{8}}dx}{x^{1-u}(x-1)^{\frac{5}{2}(u+1)}}, 
\end{split}
\end{align}
remembering that $T_\text{eff}$ also depends on the radius. Here, we rescaled the variable over which we integrate, so now $x = R_p/R_F$ and $x_0 = R_0/R_F$. Let us notice that it takes an infinite amount of time for the planet to contract fully; this result is independent of the theory of gravity.

\subsection{Numerical solutions}
\begin{figure*}[t]
\centering
\includegraphics[scale=.95]{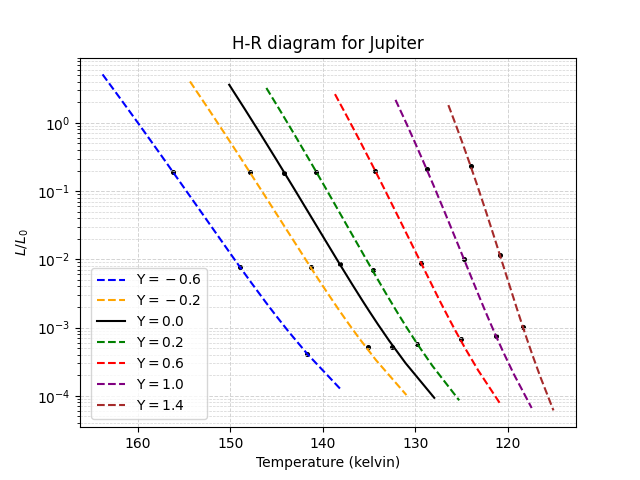}
\caption{[color online] H-R diagram for Jupiter. Different curves correspond to different values of the $\Upsilon$ parameter, characterizing deviations for GR (here, the black solid line). On the y-axis, we put the scaled luminosity of the object ($L_0 = 10^{29} erg/s$). The dots represend different moments of time (from top to bottom): $t = 10^6$ years, $t = 10^8$ years, and $t = 5\times 10^9$ years.}
\label{jup_fig}
\end{figure*}

The solution was obtained by solving numerically the equation \eqref{cond} (using the bisection method) for a range of possible radii of a planet of mass equal to Jupiter's mass. Each solution gave us a direct relationship between the effective temperature and other parameters characterizing the system, which was then used to compute the luminosity. When computing the equilibrium temperature $T_\text{eq}$, we assumed that the planet's mass was equal to Jupiter's, and also its distance from the parent star was $\approx 5$ AU. The procedure was repeated for different values of the parameter $\Upsilon$. The results of this procedure can be seen in Figure \ref{jup_fig}, where on the y-axis we put the scaled luminosity of the object ($L_0 = 10^{29} erg/s$). The black dots represent different moments of time: the uppermost is for $t = 10^6$ years, the middle one for $t = 10^8$ years, and the lowest for $t = 5\times 10^9$ years. Different times were obtained by integrating numerically the integral \eqref{age} for appropriate radii. As one can see, bigger values of the parameter $\Upsilon$ correspond to lower temperatures for the same range of the planet's radii, but also the cooling rate is slightly lower, as the final dot for the time $t = 5\times 10^9$ years lies above the line $L/L_0 = 10^{-3}$, whereas the other dots are located beneath that line. 

\begin{figure*}[t]
\centering
\includegraphics[scale=.95]{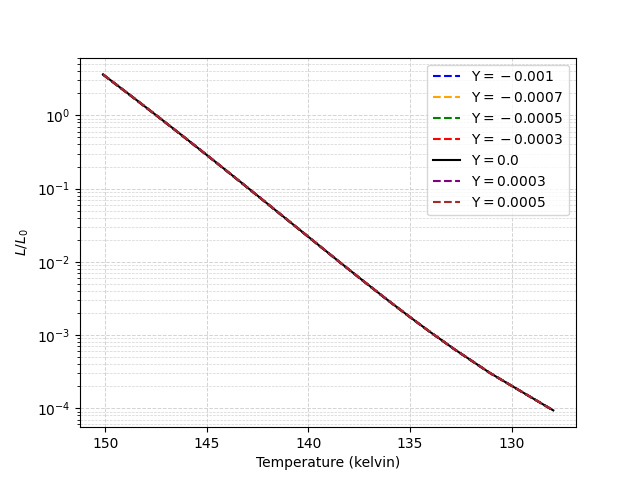}
\caption{[color online] H-R diagram for Jupiter. Different curves correspond to different values of the $\Upsilon$ parameter, characterizing deviations for GR (here, the black solid line). On the y-axis, we put the scaled luminosity of the object ($L_0 = 10^{29} erg/s$). }
\label{jup_fig_2}
\end{figure*}

{We repeated the calculations for a tighter range of the parameter $\Upsilon \in [-10^{-3}, 5\times 10^{-4}]$ to illustrate how minuscule the effect might be for very small values of $\Upsilon$ (see Fig \ref{jup_fig_2}).}

\section{Uncertainties analysis}
In this section, we obtain analytical formulas allowing us to quantify the variation in observable quantities with respect to certain parameters assumed in our calculations. Since, as obtained by our numerical analysis, we expect that the modified gravity effects are crucial in the late times of the substellar objects' evolution, we will determine whether any alternation of the rate of change of degeneracy due to the presence of Horndeski's parameter $\Upsilon$ could be overshadowed by uncertainties in other theory parameters, i.e. ionization fraction $x_{H^+}$ or different mass fractions of hydrogen and helium. In what follows, all quantities computed for $\Upsilon = 0$ and the ionization and mass fractions assumed in the part of the article preceding this section are denoted with the subscript $0$. For each quantity $Q$, $\delta Q$ is to be understood as:
\begin{equation}
\begin{split}
  & \delta Q :=  Q(\text{modified values of } x_{H^+}, X, Y; \Upsilon \neq 0)\\
   &- Q(x_{H^+}=0.255, X=0.75, Y=0.25; \Upsilon=0)\\
   &\equiv Q - Q_0. 
\end{split} 
\end{equation}
First, let us compute the relative change in the degeneracy using Eq \eqref{degeneracy}:

\begin{equation}
    \frac{\delta\psi}{\psi_0} = \frac{\delta T}{T_0} -\frac{2}{3}\frac{\delta\rho}{\rho_0}
\end{equation}
Variation in the temperature is calculated using Eq \eqref{radius0} and Eq \eqref{radius_density}:
\begin{equation}
    \frac{\delta T}{T_0} = \frac{\delta\mu}{\mu_0} + \frac{1}{3}\frac{\delta\rho}{\rho_0}
\end{equation}
so that
\begin{equation}
    \frac{\delta\psi}{\psi_0} = \frac{\delta \mu}{\mu_0} -\frac{1}{3} \frac{\delta\rho}{\rho_0}.
\end{equation}

At the end, we compute the variation in rate of change in time of the degeneracy parameter:
\begin{equation}
    \frac{\dot{(\delta\psi)}}{\dot{\psi_0}} = \frac{4\Upsilon}{7} + \frac{\delta \mu_{1mod}}{\mu_{1mod,0}} - \frac{8}{3}\frac{\delta\mu_e}{\mu_{e,0}} + \frac{12 \psi_0 \delta a}{7(1 + a_0\psi_0 + b_0)} + \frac{\delta\psi}{w_0}
\end{equation}
where $\Upsilon \ll 1 $ and:
\footnotesize{
\begin{equation}
    \begin{split}
        w_0 & = \frac{\mathcal{F}_0}{\mathcal{G}_0}
    \end{split}
\end{equation}
}
where
\begin{equation}
    \begin{split}
        \mathcal{F}_0 = & 21\left(1 + e^{\frac{1}{\psi_0}}\right)\psi_0\Big(2(8 + 8a_0\psi_0+5\pi^2 \psi_0^2 \\
        &- 5\psi_0 \text{Log} \left(1 + e^{-\frac{1}{\psi_0}}\right) + 30\psi^2 \text{Li}_2\left(-e^{-\frac{1}{\psi_0}}\right) \Big)
    \end{split}
\end{equation}
and
\begin{equation}
    \begin{split}
        \mathcal{G}_0 = & 2\Big(-10 + 80e^{\frac{1}{\psi_0}} + \left(1 + e^{\frac{1}{\psi_0}}\right)\big(368 a_0\psi_0 + 410\pi^2\psi_0^2 \\&- 655\psi_0 \text{Log}\left(1 + e^{-\frac{1}{\psi_0}}\right) + 1230\psi_0^2\text{Li}_2\left( -e^{-\frac{1}{\psi_0}}\right)\big)\Big)
    \end{split}
\end{equation}

\begin{figure}[h!]
\flushleft 
\advance\leftskip-1cm
\includegraphics[scale=.44]{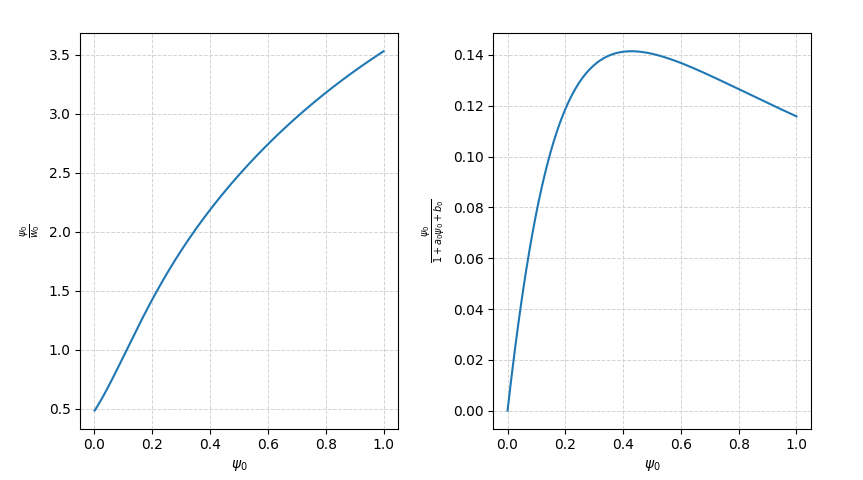}
\caption{[color online] Change of $\frac{\psi_0}{w_0}$ and $\frac{\psi_0}{1+a_0 \psi_0 + b_0}$ parameters with respect to degeneracy value. The numerical value of $a_0$ is $a_0 = 2.8679$. The value $\Psi_0=1$ denotes the non-degenerated case (early times), see the figures \ref{deg1} and \ref{deg2}.}
\label{w}
\end{figure}

\normalsize
Taking into account the formulae above and the definition of $a_0$, we can write:
\begin{equation}
    \frac{\delta a}{a_0} = -\frac{\delta \mu_1}{\mu_{1,0}}
\end{equation}
and, finally:
\begin{equation}
\begin{split}
    \frac{\dot{(\delta\psi)}}{\dot{\psi_0}} = & \frac{4\Upsilon}{7} + \frac{\delta \mu_{1mod}}{\mu_{1mod,0}}- \frac{8}{3}\frac{\delta\mu_e}{\mu_{e,0}}  - \frac{12 \psi_0 \delta \mu_1}{7\mu_{1,0}(1 + a_0\psi_0 + b_0)}\\
    &+ \frac{\psi_0}{w_0}\left(\frac{\delta \mu}{\mu_0}-\frac{1}{3} \frac{\delta\rho}{\rho_0}\right).
\end{split}  
\end{equation}
The density variation for $r \approx R$ takes the following form (expanded around $r = 0.99 R$) \cite{saltas,debora}:

\begin{equation}
    \frac{\delta \rho}{\rho_0} = -0.21 \Upsilon - 10.1\Upsilon\left(\frac{r-R_{cz}}{R}\right)
\end{equation}
where $R_{cz}$ is the radius of the convective zone. Putting this altogether, we get:
{\footnotesize
\begin{equation}
\begin{split}
    \frac{\dot{(\delta\psi)}}{\dot{\psi_0}} &= \left(\frac{4}{7}+\frac{\psi_0\left(0.21 + 10.1\left(\frac{r-R_{cz}}{R}\right)\right)}{3 w_0}\right)\Upsilon + \frac{\delta \mu_{1mod}}{\mu_{1mod,0}} - \frac{8}{3}\frac{\delta\mu_e}{\mu_{e,0}} \\
    &- \frac{12 \psi_0 \delta \mu_1}{7\mu_{1,0}(1 + a_0\psi_0 + b_0)} + \frac{\psi_0}{w_0}\frac{\delta \mu}{\mu_0}.
\end{split}
\end{equation}
}
In Figure \ref{w}, we present how $\frac{\psi_0}{w_0}$ changes as a function of the degeneracy parameter. We can see that this parameter decreases as time progresses (we remind that the value of the degeneracy parameter goes down with time), going to the constant value of $\sim\frac{1}{2}$. In Fig \ref{w} we also plotted the dependence of the parameter multiplying $\delta\mu_1$; as one can see, at a certain moment it reaches maximum value but then drops to zero at later times. 

For large times and $r=R$, we obtain the following:
\begin{equation}
\begin{split}
    \frac{\dot{(\delta\psi)}}{\dot{\psi_0}} & \approx  0.727\Upsilon + \frac{\delta \mu_{1mod}}{\mu_{1mod,0}} - \frac{8}{3}\frac{\delta\mu_e}{\mu_{e,0}} + \frac{1}{2}\frac{\delta \mu}{\mu_0}  \\
    & =  0.727\Upsilon + \frac{\delta \mu_{1mod}}{\mu_{1mod,0}}- \frac{13}{6}\frac{\delta\mu_e}{\mu_{e,0}},
\end{split}  
\end{equation}
where in the last equality we have used the assumption on the zero metallicity, so $\mu=\mu_e$.

\begin{center}
    \textit{Varying with respect to $x_{H^+}$}
\end{center}

\begin{figure}[h!]
\includegraphics[width=\linewidth]{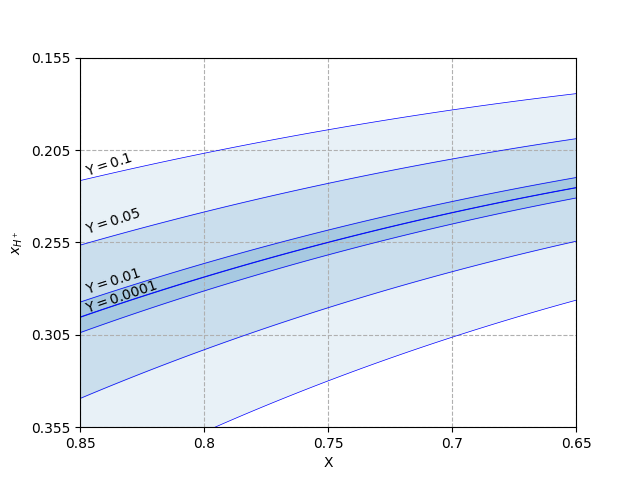}
\caption{[color online] Regions of possible values $x_{H^+}$ and $X$ (and, consequently, in $Y$) resulting in smaller joint modifications than coming from variations in $\Upsilon$ alone. }
\label{regions}
\end{figure}

In this part, we want to investigate how a simultaneous change in $x_{H^+}$ and $X$ will influence the evolution of the electron degeneracy. We neglect the possible change in other theory parameters, such as $b_1$ and $\nu$, since in Fig \ref{regions} we plot regions of values of deviations from $X=0.75, Y=0.25, x_{H^+}=0.255$, resulting in smaller joint variations than the ones coming from modified gravity alone, i.e.
\begin{equation}
    \begin{split}
        \left|\frac{\dot{(\delta\psi)}}{\dot{\psi_0}} (X=0.75, Y=0.25, x_{H^+}=0.255; \Upsilon \neq 0)\right| \\
         > \left|\frac{\dot{(\delta\psi)}}{\dot{\psi_0}}(X, Y, x_{H^+}; \Upsilon = 0)\right|.
    \end{split}
\end{equation}
As one can see, the regions shrink down to a single line for very small values of the parameter $\Upsilon$, representing possible values of $X$ and $x_{H^+}$ resulting in $\frac{\delta \mu_{1mod}}{\mu_{1mod,0}} - \frac{13}{6}\frac{\delta\mu_e}{\mu_{e,0}} = 0$. We impose additional constraints on $X$ and $x_{H^+}$, coming from theories considered in \cite{maria}.

To get some idea about the order of variations coming from either modified gravity or changed parameters $X$ and $x_{H^+}$, let us compute the value $\left|\frac{\dot{(\delta\psi)}}{\dot{\psi_0}}\right|$ for extreme values of these parameters:
\begin{table}[h!]
\centering
\begin{tabular}{ |c||c|c|c| } 
 \hline
   & $x_{H^+} = 0.105$ & $x_{H^+} = 0.255$ & $x_{H^+} = 0.26$ \\
  \hline\hline
 $X=0.7$ & $0.20$ & $0.02$ & $0.02$ \\ 
 $X=0.75$ & $0.21$ & $0$ & $0.01$ \\ 
 $X=0.8$ & $0.22$ & $0.02$ & $0.01$ \\ 
 \hline
\end{tabular} 
\caption{$\left|\frac{\dot{(\delta\psi)}}{\dot{\psi_0}}\right|$ values for different combinations of $X$ and $x_{H^+}$ when no modified gravity effects are taken into account.}
\end{table}\\

On the other hand, $\left|\frac{\dot{(\delta\psi)}}{\dot{\psi_0}}\right|$ for unchanged $X$ and $x_{H^+}$ and $\Upsilon \in \{0.1, 0.05, 0.01, 0.0001\}$ are, respectively: $0.072, 0.036, 0.007$, $0.000007$. 

\begin{center}
    \textit{Keeping $x_{H^+}$ constant}
\end{center}

\begin{figure}[h!]
\includegraphics[width=\linewidth]{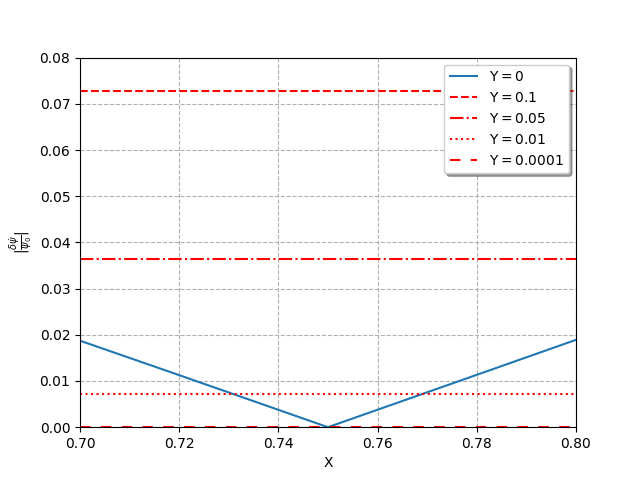}
\caption{[color online] Comparison of the uncertainties resulting from varying the hydrogen mass fraction and from modified gravity while keeping the ionization fraction $x_{H^+}$ fixed.}
\label{errors_x}
\end{figure}

We want to repeat the procedure outlined above while keeping the ionization fraction $x_{H^+}$ constant. This corresponds to a situation when we decide on a given model without assuming any uncertainties in $x_{H^+}$; in the case of our paper, we decided to choose Model D with $b_1 = 2$ and $\nu = 1.6$. The only parameters we will vary in that case are: the DHOST parameter $\Upsilon$ and $X$, the hydrogen mass fraction (one is reminded that $Y = 1 - X$). Again we compare absolute joint contributions to $\left|\frac{\dot{(\delta\psi)}}{\dot{\psi_0}}\right|$ when $\Upsilon=0$ and $\Upsilon\neq 0$. We present the results in Fig \ref{errors_x}. In this figure, the solid blue line corresponds to the situation when we vary $X$ but keep $\Upsilon =0$; the dashed and dotted lines represent uncertainties coming from the modification of $\Upsilon$ value only, with $X = 0.75$ in all four cases. The figure allows us to determine for what values of $X$, the effects coming from modifications of a star's composition are greater than modifications coming from extended gravity alone. 

\section{Discussion and conclusions}

In this work, we have focused on the cooling processes of brown dwarf stars and Jupiter-like planets in the framework of {DHOST} theories of gravity. Those theories modify the hydrostatic equilibrium equation which is used to obtain the equations ruling cooling models. Because of that fact, we had expected some differences with respect to the results which are based on Newtonian gravity. 

In the case of brown dwarfs, we have used a realistic analytical equation of state describing the complex interior of those objects: the main issue is related to taking into account the phase transition between the molecular hydrogen in the photosphere and the ionized one, which is present in the brown dwarfs' interior, according to the current models \cite{aud,chab}. Moreover, the considered EoS suits well when a mixture of degenerate Fermi gas with hydrogen's and helium's ions must be considered which is the case of the brown dwarfs' interiors. This allowed us not only to derive the master equations describing the inner structure of those objects, but also to find the photospheric quantities which are crucial for modelling cooling processes, \eqref{fotpres}, \eqref{tefflum}, and \eqref{lum}, being dependent on the varying electron degeneracy \eqref{deg_ev}. 

The solutions of the equations governing the cooling process in brown dwarfs are given by the figures \ref{deg1}, \ref{deg2}, and \ref{tlum}. For the given values of the parameter $\Upsilon$ we see that for an older object one deals with the bigger differences with respect to the Newtonian model. It is, as expected, more evident for the higher absolute values of the parameter. For example, for the $1$ Myr brown dwarf, the difference is about $5\%$ while for the $10$ Gyr one it differs by $10-15\%$ for $\Upsilon=|0.6|$. In the case of the luminosity, the ratios are more significant: $20-40\%$ for old brown dwarfs, while for the young ones we deal with $10-15\%$. Similarly, as in the case of the previous results in the field \cite{maria}, scalar-tensor theories could be constrained by future data. 

{Since the positive values of $\Upsilon$, as long as the second derivative of mass with respect to the radius is negative, correspond to weaker gravity (and vice versa), brown dwarfs in such models will be smaller (bigger), as it also follows from Eqs \eqref{radius}-\eqref{central density}. Also, their temperature at the core is lower as the theory parameter increases. The situation becomes somewhat counter-intuitive when one considers the photosphere of a brown dwarf. It becomes apparent from Figures \ref{deg1} and \ref{tlum} that, as the gravity becomes weaker, the luminosity drops faster, while the opposite effect is visible for the degeneracy (one is reminded that $\Psi = 1$ corresponds to no degeneracy). The interpretation is clear: for weaker gravity, the stored energy can be radiated faster, therefore it has the same effect on the cooling process. Let us recall that the electron degeneracy is a response (via the Pauli exclusion principle) to the attractive nature of gravity in the interiors of the substellar objects, such that weakening this interaction will also have a direct effect on the degeneracy. We see from the eq. \eqref{deg_ev} that the positive (negative) range of the parameter makes the matter reach a more degenerate state in a shorter (longer) time with respect to the Newtonian case.}

On the other hand, although gaseous giant planets are quite similar to brown dwarfs, the methods allowing to obtain the cooling model are much more complicated. The main and the most important difference is that one cannot neglect the energy source coming from the parent star {if we do not deal with a rogue planet} (let us notice that we neglected tidal and ohmic heating in our analysis). It results in slightly different atmospheric characteristics, that is, the effective temperature depends on the equilibrium one, which is a property of a given planetary system and its parent star (see the equations \eqref{teq} and \eqref{temp}). Because of that fact, we followed a simplified approach developed in \cite{don,anetaJup}, whose main assumption is related to the simplified description of the matter behaviour: that is, one models it as a slightly modified polytropic equation of state \eqref{p2}. However, even with such a toy model we were able to get solutions close to the realistic values for the considered theories of gravity, and again, as expected, the evolutionary paths are affected with respect to Newtonian gravity \ref{jup_fig}. Depending on the $\Upsilon$'s sign, Jupiter's age, based on its current effective temperature, can be very different from the one we believe it is. For instance, the age and average effective temperature of Jupiter is $\approx4.9\times10^9$ years and $\approx130$ K, respectively, according to the Newtonian model. {Our calculations reveal that for the broader range of the parameter $\Upsilon$, the age of Jupiter varies from $2.08\times t_J$ for $\Upsilon = -0.2$ to $8.16\times 10^{-6} \times t_J$ for $\Upsilon = 1$ (smaller and larger values of the $\Upsilon$ were beyond our integration interval, but the tendency is pretty clear), where $t_J$ is the age of Jupiter contracting from the initial size of $R_0 = 10^{12} \text{m}$. For the much tighter range of the parameter $\Upsilon$, we got $0.9993\times t_J$ for $\Upsilon = -10^{-3}$, and $1.0007\times t_J$ for $\Upsilon = 5\times10^{-4}$. It must be noted that for the smaller values of the $\Upsilon$, uncertainties coming from the temperature estimates are much higher than the ones introduced by the modification of gravity.} {The modification introduced by DHOST theories slows down the process of cooling down of jovian planets. They also achieve lower temperatures for greater values of $\Upsilon$. At a given value of the temperature, the objects have bigger luminosity for greater $\Upsilon$. Also, for a given luminosity, the objects become older as $\Upsilon$ increases.} This also means that the evolution of our Solar System will be also distinct from the one we are used to. 

{
Let us also notice that the giant planets' models presented in this work cannot be used yet to constrain theories of gravity. This is so according to the fact that in order to derive the cooling equations we have used approximated, analytical form of the equations of state, and, more importantly, a very simplified atmosphere description, mainly related to the opacity models. Apart from this, some of those microscopic properties can depend on a theory of gravity \cite{kim,del,hos1,hos2,awmicro,kalitaspecific} and therefore, should be also properly re-analyzed before applying them to the stellar and substellar modelling, and finally, to constrain models of gravity. The problem is however different in the case of brown dwarfs' modelling, as it reveals our uncertainties analysis.
}

 { Regarding the uncertainties, we have demonstrated that the evolution of the electron degeneracy \eqref{deg_ev} is a crucial element in modelling substellar objects. It depends on the composition, temperature, energy density, opacity, ionization, and phase transition points. As demonstrated in \cite{debora}, the atmosphere modelling carries the highest uncertainties mainly related to metallicity in the case of low-mass stars. Since in our case we have neglected metallicity and considered a simple Rossland opacity (which in the general case depends on energy density and composition), we do not have such a dependence. We are aware that metallicity plays a very important role in the substellar evolution \cite{spiegel}. Hence, the analysis in that direction will have to be done in the near future.}
 
 {
 From our current uncertainty analysis, in the case of fixed ionization fraction $x_{H^+}$, presented in Fig \ref{errors_x}, we can see that the effect of modified gravity is more pronounced in the admissible range of $X$ if the modifications are of order $\geq 10^{-2}$ (for $\Upsilon = 10^{-2}$, uncertainties coming from $X$ can be greater, but remain of the same order). The value of $10^{-4}$ however rules out any possibility to produce a noticeable effect compared to a varied $X$, and thence such modification will not have any noticeable effect on the evolution of the electron degeneracy. }

{
 However, taking into account more variables, such as small changes in the ionization, as plotted in Fig \ref{regions}, for each value of the $\Upsilon$ parameter there exists a region of possible values of $x_{H^+}$ and $X$ such that their joint effect on the rate of change of degeneracy is smaller than the one coming from a modification of gravity. As expected, the smaller value of the parameter $\Upsilon$, the smaller region we deal with; nevertheless, the effect is still present even for the very restrictive range of the parameter \cite{saltas}. Since the luminosity and effective temperature depend on the time evolution of the electron degeneracy and the effects of modified gravity accumulate with time \cite{suppl} (compare also the ratios in the figure \ref{tlum} and \ref{deg1}), one deals with with a possibility to test this theory with the brown dwarf stars.
 }

{Nevertheless, this is the first step undertaken in the scalar-tensor theories to understand how to use the interior properties of planets and brown dwarf stars to test such theories, as mentioned in the introduction. The effects of the extra terms appearing in our equations can be interpreted as additional heating or cooling processes, resulting in altered evolutionary scenarios with respect to the Newtonian model. Even improved models of brown dwarfs and jovian planets (for example, the next step could be considering rotation effects on the electron degeneracy evolution \cite{chowdROT}) will carry uncertainties related to theoretical assumptions - however, having a large sample of observational properties, differentiated with respect to distances from the detectors, types of the objects, and the objects' neighbourhood, will allow to reduce the error, and finally, to constrain the models \cite{smirnov,suppl,gaia}. Moreover, knowing the age of the neighbourhood structures of giant planets and brown dwarfs\footnote{Notice that some age determination techniques depend on a model of gravity, too \cite{aneta3}.} which are expected to form at the same time as the substellar bodies, could also be used to test theories, since the age is the most affected quantity by modified gravity. Data on the Solar System objects, because of the vicinity ensuring higher accuracy, such as effective temperatures (which can be also derived from theoretical models), measurements of the energy flux radiated from the Sun and received by a planet as well as radiated away from it, together with seismic data providing information on internal properties are only a part of the opportunities which can be used to test theories. So far, there have been just a few works discussing those possibilities and it is expected that there will appear more in the nearest future.
}

We will leave further considerations along these lines for future work. However, we should again underline that if we believe that there is a bit better theory of gravity, allowing to describe gravitational phenomena on a much wider scale than GR, probably that theory will also slightly modify the Newtonian limit. 
Research in this direction is in high demand, especially in light of many current and future missions, whose aim is to explore our and other planetary systems and to provide more accurate data regarding the substellar objects \cite{vision,voyage,webb,nancy,tess,spitzer,nn}.

\vspace{5mm}
\noindent \textbf{Acknowledgement.} 
This work was supported by the EU through the European Regional Development Fund CoE program TK133 ``The Dark Side of the Universe." AW acknowledges financial support from MICINN (Spain) {\it Ayuda Juan de la Cierva - incorporac\'ion} 2020 No. IJC2020-044751-I.


\end{document}